\documentclass[%
 reprint,
 superscriptaddress,
 amsmath,amssymb,
 aps,
]{revtex4-1}

\usepackage{graphicx}% Include figure files
\usepackage{dcolumn}% Align table columns on decimal point
\usepackage{bm}% bold math
\usepackage{soul}
\usepackage{color}
\usepackage{physics}
\usepackage{subfig}
\usepackage{natbib}
\usepackage{booktabs}
\usepackage{float}
\usepackage{caption}
\newcommand{\e}{\text{e}}

\begin{document}

\preprint{APS/123-QED}

\title{Measurement-device-independent quantum key distribution with asymmetric sources}% Force line breaks with \\
	\author{Jia-Ju Deng}\email{These authors contributed equally to this work}
	\affiliation{CAS Key Laboratory of Quantum Information, University of Science and Technology of China, Hefei, Anhui 230026, P. R. China}
	\affiliation{CAS Center for Excellence in Quantum Information and Quantum Physics, University of Science and Technology of China, Hefei, Anhui 230026, P. R. China}
	\author{Feng-Yu Lu}\email{These authors contributed equally to this work}
	\author{Zhen-Qiu Zhong}
    \author{Xiao-Hai Zhan}
	\author{Zhen-Qiang Yin}
	\email{yinzq@ustc.edu.cn}
	\author{Shuang Wang}
	\email{wshuang@ustc.edu.cn}
	\author{Wei Chen}
	\author{De-Yong He}
	\author{Guang-Can Guo}
	\author{Zheng-Fu Han}
	\affiliation{CAS Key Laboratory of Quantum Information, University of Science and Technology of China, Hefei, Anhui 230026, P. R. China}
	\affiliation{CAS Center for Excellence in Quantum Information and Quantum Physics, University of Science and Technology of China, Hefei, Anhui 230026, P. R. China}
	\affiliation{Hefei National Laboratory, University of Science and Technology of China, Hefei 230088, China}
% \author{Feng-Yu Lu}
%  \altaffiliation[Also at ]{Physics Department, XYZ University.}%Lines break automatically or can be forced with \\
% \author{Zheng-Qiang Yin}%
%  \email{Second.Author@institution.edu}
% \affiliation{%
%  Authors' institution and/or address\\
%  This line break forced with \textbackslash\textbackslash
% }%

% \collaboration{MUSO Collaboration}%\noaffiliation

%  \homepage{http://www.Second.institution.edu/~Charlie.Author}
% \affiliation{
%  Second institution and/or address\\
%  This line break forced% with \\
% }%
% \affiliation{
%  Third institution, the second for Charlie Author
% \affiliation{%
%  Authors' institution and/or address\\
%  This line break forced with \textbackslash\textbackslash
% }%

% \collaboration{CLEO Collaboration}%\noaffiliation

\date{\today}% It is always \today, today,

\begin{abstract}
	Measurement-device-independent quantum key distribution (MDI-QKD), which eliminates all the attacks from the eavesdropper to the measurement party, has been one of the most promising technology for the implementation of end-to-end quantum networks.
	In practice, the asymmetry of both sources and channels is generally inevitable.
	Therefore, we propose a theory to analyze the performance when any two MDI users in networks communicates using asymmetric sources in distinct single or multiple temporal modes.
	As a specific application, we model to obtain the key rate of MDI-QKD with weak coherent pulse source and spontaneous parametric down-conversion source, and compare the performance to the cases with symmetric (i.e. identical) sources.
	The result demonstrates that the actual performance does not degrade due to the asymmetry of sources.
	In contrary, it maintains at a good level over the entire distance we study.
	This work provides a theoretical basis for analyzing and optimizing MDI-QKD networks with asymmetric sources, and thus paving the way for the practical deployment of completely asymmetric MDI-QKD networks.
\end{abstract}

\pacs{Valid PACS appear here}% PACS, the Physics and Astronomy
\maketitle

%\tableofcontents

\section{Introduction}
Quantum key distribution (QKD) \cite{QKD1,QKD2,QKD3} has achieved enormous development in the past 40 years.
The inherent security of QKD originates from the principle of quantum mechanics \cite{LoUnconditional, PirandolaAdvanced, Renner, Scarani}, and thus it allows two remote users, Alice and Bob, to exchange secret keys with information-theoretical security.
Moreover, QKD is the fundamental step towards the future quantum Internet \cite{kimble2008quantum,castelvecchi2018quantum,wehner2018quantum,singh2021quantum,li2023entanglement}.

Although this technology is theoretically secure, practical implementations of QKD inherently suffer from device imperfections, which lead to potential security risks.
Measurement-device-independent (MDI) QKD \cite{MDI} allows an untrusted third party to perform measurements and announces the results, which eliminates all the attacks from the eavesdropper to the measurement party \cite{Braunstein}.
MDI-QKD is mature and easy to implement, so it has been worldwide studied with numerous remarkable achievements in theory and experiment
\cite{Curty,WangAsymmetric,Ma,wang2013three,ma2012alternative,xu2013practical,yu2015statistical,zhou2016making,lu2020efficient,liu2013experimental,tang2014measurement,pirandola2015high,wang2015phase,
tang2016measurement,comandar2016quantum,yin2016measurement,roberts2017experimental,wang2017measurement,liu2018polarization,semenenko2020chip,wei2020high,woodward2021gigahertz,zhou2021reference,lu2022unbalanced,lu2023hacking,erkilicc2023surpassing}.
Particularly, in MDI-QKD networks, the star topology is quite suitable, in which the measurement unit locates at the central relay \cite{TangMetro,fan2021measurement,FanYuan}.
Therefore, MDI-QKD is a promising scheme for metropolitan networks \cite{ChenIntegrated, TangMetro, Tang}, and simultaneously an achievable and specific solution for upgrading the trusted repeater quantum network \cite{wehner2018quantum, fan2021measurement}.

\begin{figure}[h]
	\centering
	\includegraphics[width=\columnwidth]{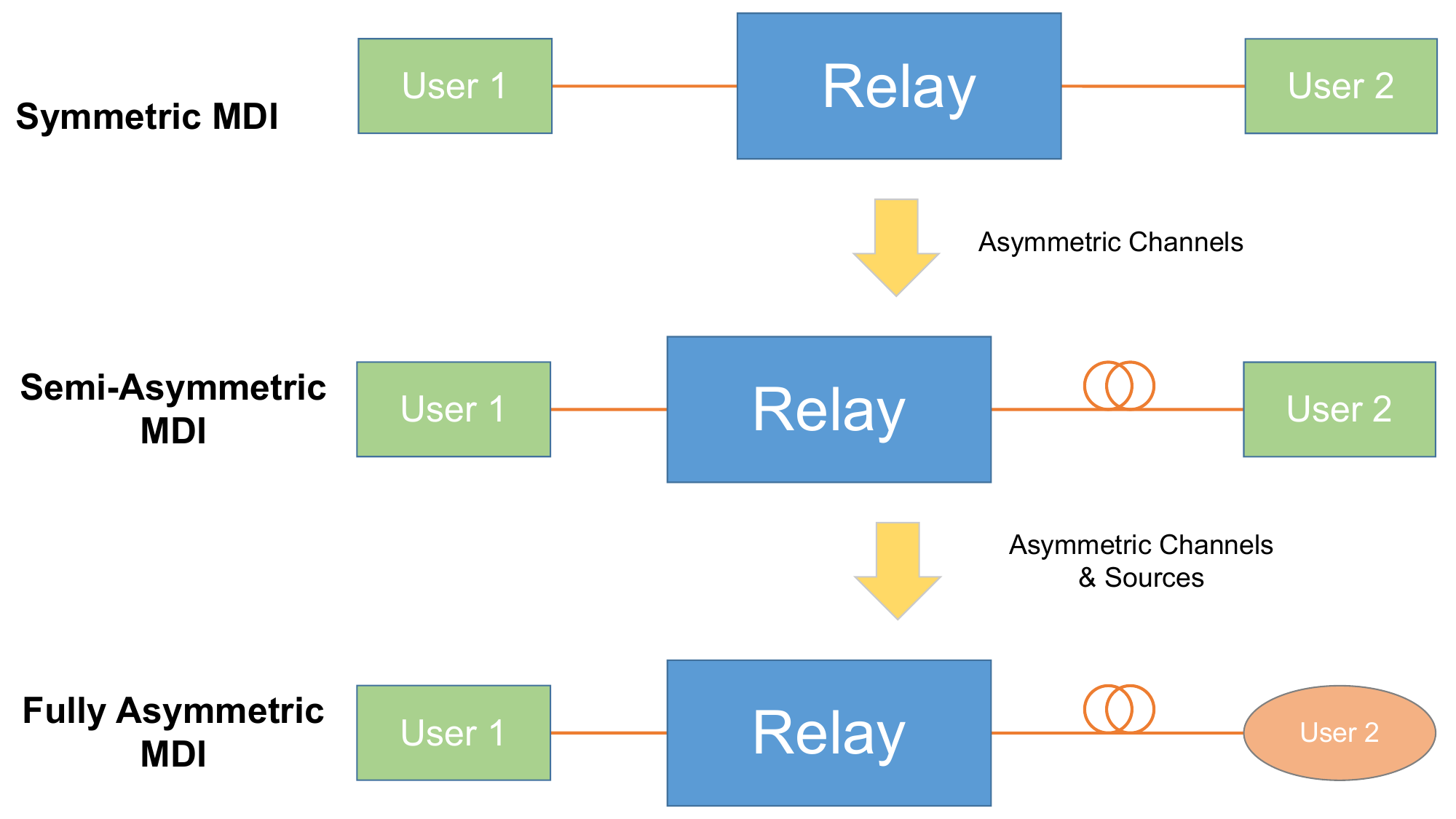}
	\caption{A Scheme of the evolution of the asymmetric MDI-QKD.
	The structure at the top is the typical symmetric MDI-QKD, in which two users the identical sources and distribute at equal distances.
	The next structure shows that the distances from two users to the relay differ.
	Now we take this asymmetry of MDI-QKD a step further: we consider the sources of two users are symmetric.
	This is what the structure at the bottom demonstrates.}
\end{figure}

However, these studies share a common limitation that they assume symmetric channels and symmetric (i.e. identical) sources.
In practical networks, it is unreasonable and impossible to require all users to distribute symmetrically at equal distances around the relay, and inconvenient for network users to employ customized transmitters with the same parameters.
The countermeasure against asymmetry channels has been solved by separately choosing intensities to compensate for different channel loss \cite{WangAsymmetric,lu2019parameter,liu2019experimental}.
Nevertheless, research on asymmetric source communication, that is, communication using sources with distinct waveforms in frequency domain, is still lacking.
In previous work on quantum interference \cite{Agata,Jin,TakesueInterference,Tichye,Young}, the case of multiple photons is ignored or avoided.
And their theories only give the two-fold coincidence probability, without considering the polarization degree of freedom.
Simultaneously, it is challenging to extend the theoretical framework to include such information.
Besides, how distinct multiple temporal modes (see appendix \ref{appendix: SPDC} for specific concept) match is still an unanswered question.
All these problems are obstacles to the construction of the MDI-QKD model with asymmetric sources.

Now in this paper, we propose a new theory to analyze how two parties communicate with asymmetric sources in the MDI-QKD protocol.
This theory studies the interference of multiple photons in different temporal modes on beam splitter (BS).
In this process, polarization information can be naturally contained.
Generally, for two arbitrary light fields with given wave functions in time or frequency domain, it can be used to obtain the final state after BS.
In principle, this advantage allows us to perform the theoretical analysis and optimization of the lower bound of the key rate in MDI-QKD with any asymmetric sources.
Furthermore, we decouple the intensity and the wave function of a practical signal, so the method of enhancing the key rate with asymmetric channels proposed in Ref. \cite{WangAsymmetric} is still effective.
It is noted that the theoretical framework we proposed can provide valuable theoretical fundamental for implementing the practical large-scale quantum network, in which users usually hold diverse sources in communications.
Based on this, we can experimentally optimize the parameter settings to attain the key rate nearly as high as the simulation result.

This paper is structured as follows.
In Sec. \ref{sec: multi-mode interference}, we first introduce the theory of the interference of two single photons in different single modes, then we extend the method to multimode case, and give several examples as verification.
In Sec. \ref{sec: asymmetric MDI}, as a specific application, we consider the communication scenario where one party use a weak coherent pulse (WCP) source and the other uses a spontaneous parametric down-conversion (SPDC) source.
We start from a simplified scheme of the experiment setup, then we give some details in this protocol and the simulation result, the curves of key rate versus distance and data size.
Simultaneously, we fix the detection efficiency and dark count at the measurement party, and compare this result with the key rates when both parties use identical WCP sources and use identical SPDC sources.
Finally, we construct a hypothetical case where the experiment parameters are attempted without the theoretical guidance.
We arbitrarily choose several fixed parameters to compare the performance of this system with the optimized one.

\section{Interference Theory of Multiple Temporal Modes}
\label{sec: multi-mode interference}
Although there has been extensive studies \cite{Agata,Jin,TakesueInterference,Tichye,Young} on the Hong-Ou-Mandel interference \cite{HOM} between a pair of photons in different modes on the 50:50 BS,
these studies share a common limitation that they consider only a single photon on a path, or even a single mode.
Therefore, we propose a new general quantum interference theory to calculate the output results after BS.
% The significant advantage of this theory is that we can clearly obtain the state after interference containing the polarization information.
% \textcolor{blue}{The polarization compatibility} allows us to truly study the details of asymmetric source communications in different modes in the framework of MDI-QKD \cite{WangMDI}.

In this section, we first introduce from the single-mode case, and then extend the conclusion to the multimode case.
Finally, we simply verify this theory through several examples with experimental evidence.

\subsection{Single-Mode Interference}
\label{subsec: independence of different TMs}
First, we define the creation operator in the temporal mode (TM) $\psi(\omega)$ \cite{Agata,Brecht}:
\begin{align}
	\hat{a}_\psi^\dag &= \int\dd\omega\psi(\omega)\hat{a}^\dag(\omega),
\end{align}
where the wave function in frequency domain $\psi(\omega)$ is probability normalized, i.e. $\int\dd\omega\abs{\psi(\omega)}^2=1$.
See appendix \ref{appendix: SPDC} for the specific concept of "temporal modes".
This expression can be seen as the coherent superposition of the single photon state around the frequency $\omega$.
In other words, a single photon can be “create” around the frequency $\omega$ with the probability $\abs{\psi(\omega)}^2\dd{\omega}$.

Now we consider a creation operator in mode $\varphi$.
Generally, $\hat{a}_\psi^\dag$ and $\hat{a}_\varphi^\dag$ do not commute.
Similar to the method used in Ref. \cite{Tichye,Young}, we define
\begin{align}
	c &= \braket{1,\psi}{1,\varphi}=\mel{0}{\hat{a}_\psi\hat{a}_\varphi^\dag}{0}\notag\\
	&= \int\dd\omega\int\dd\omega'\psi^*(\omega)\varphi(\omega')\mel{0}{\hat{a}(\omega)\hat{a}^\dag(\omega')}{0}\notag\\
	&= \int\dd\omega\psi^*(\omega)\varphi(\omega)
\end{align}
as the description of their temporal overlap, formulated using Eq.\eqref{eq: commutative relation}.
Similarly, it can be calculated that
\begin{align}
	[\hat{a}_\psi,\hat{a}_\varphi^\dag] &= \int\dd\omega\int\dd\omega'\psi^*(\omega)\varphi(\omega')[\hat{a}(\omega),\hat{a}^\dag(\omega')]\notag\\
	&= \int\dd\omega\psi^*(\omega)\varphi(\omega)=c.
\end{align}
Now we can decompose $\hat{a}_\varphi^\dag$ into components “parallel” and “orthogonal” to $\hat{a}_\psi$, that is,
\begin{align}
	\hat{a}_\varphi^\dag=c\hat{a}_\psi^\dag+\sqrt{1-\abs{c}^2}\hat{a}_\bot^\dag.
\end{align}
We prove the following three properties of the “orthogonal” component $\hat{a}_\bot^\dag$ below.
\begin{enumerate}
	\item \textbf{Commutativity} $\hat{a}_\psi$ and $\hat{a}_\bot^\dag$ commute:
	\begin{align}
		[\hat{a}_\psi,\hat{a}_\bot^\dag] &= (1-\abs{c}^2)^{-1/2}[\hat{a}_\psi,\hat{a}_\varphi^\dag-c\hat{a}_\psi^\dag]\notag\\
		&= (1-\abs{c}^2)^{-1/2}(c-c\cdot 1)=0.
		\label{eq: commutativity of different TMs}
	\end{align}
	\item \textbf{Orthogonality} $\hat{a}_\bot^\dag\ket{0}$ is orthogonal to $\hat{a}_\psi^\dag\ket{0}$:
	\begin{align}
		\braket{1,\psi}{1,\bot} &= (1-\abs{c}^2)^{-1/2}\mel{0}{\hat{a}_\psi(\hat{a}_\varphi^\dag-c\hat{a}_\psi^\dag)}{0}\notag\\
		&= (1-\abs{c}^2)^{-1/2}(c-c\cdot 1)=0.
	\end{align}
	\item \textbf{Canonical commutation relation}
	\begin{align}
		&[\hat{a}_\bot,\hat{a}_\bot^\dag]=(1-\abs{c}^2)^{-1}[\hat{a}_\varphi-c^*\hat{a}_\psi,\hat{a}_\varphi^\dag-c\hat{a}_\psi^\dag]\notag\\
		=&(1-\abs{c}^2)^{-1}(1-c^*\cdot c-c\cdot c^*+\abs{c}^2\cdot 1)=1.
	\end{align}
\end{enumerate}
Based on these three properties above, we conclude that operators $\hat{a}_\psi^\dag$ and $\hat{a}_\bot^\dag$ are creation operators in distinct Hilbert spaces.
This conclusion explains why the states obtained after TM decomposition are product states.

\subsection{Multimode Interference}
In this subsection, We generalize the distinct single-mode decomposition method from the previous subsection to the multimode case.
Let $\{\psi_n(\omega)\}_{n=1}^N$ and $\{\varphi_n(\omega)\}_{n=1}^N$ be two sets of orthonormal function bases, and
\begin{align}
	\hat{A}_{\vb*{\psi}}^\dag &= \sum_{\vb*{n}=0}^\infty\alpha_{\vb*{n}}\prod_{i=1}^N\qty(\hat{a}_{\psi,i}^\dag)^{n_i},
	\label{eq: Multimode base}\\
	\hat{A}_{\vb*{\varphi}}^\dag &= \sum_{\vb*{n}=0}^\infty\beta_{\vb*{n}}\prod_{i=1}^N\qty(\hat{a}_{\varphi,i}^\dag)^{n_i}
\end{align}
are two multimode creation operators, where
\begin{align}
	\hat{a}_{\psi,i}^\dag &= \int\dd{\omega}\psi_i(\omega)\hat{a}^\dag(\omega),\\
	\hat{a}_{\varphi,i}^\dag &= \int\dd{\omega}\varphi_i(\omega)\hat{a}^\dag(\omega)
\end{align}
are the creation operators of each mode;
$\vb*{n}=(n_1,\dotsb,n_N)$ means there are $n_i$ photons in the $i$th mode;
$\vb*{\psi}$ and $\vb*{\varphi}$ respectively represent the set $\{\psi_n(\omega)\}_{n=1}^N$ and $\{\varphi_n(\omega)\}_{n=1}^N$.

Note that the so-called “single-mode decomposition” actually involves two TMs, $\psi$ and $\bot$,
then the approach is to express each basis function in set $\{\varphi_n(\omega)\}_{n=1}^N$ as a linear combination of the basis functions in another set $\{\psi_n(\omega)\}_{n=1}^N$,
that is, to connect the two sets of basis functions via a unitary transformation matrix $\vb*{c}$: $\vb*{\varphi}(\omega)=\vb*{c}\vb*{\psi}(\omega)$.
Now we can rewrite $\hat{A}_{\vb*{\varphi}}^\dag$:
\begin{align}
	\hat{A}_{\vb*{\varphi}}^\dag &= \sum_{\vb*{n}=0}^\infty\beta_{\vb*{n}}\prod_{i=1}^N\qty(\sum_{j=1}^Nc_{ij}\hat{a}_{\psi,j}^\dag)^{n_i}\notag\\
	&= \sum_{\vb*{n}=0}^\infty\beta_{\vb*{n}}\prod_{i=1}^Nn_i!\sum_{\norm{\vb*{m}_i}_1=n_i}\prod_{j=1}^N\frac{(c_{ij}\hat{a}_{\psi,j}^\dag)^{m_{ij}}}{m_{ij}!},
	\label{eq: Multimode decomposition}
\end{align}
where multinomial theorem and Eq.\eqref{eq: commutative relation} has been applied in the last line, and now $\hat{A}_{\vb*{\varphi}}^\dag$ can be expressed as the form similar to Eq.\eqref{eq: Multimode base}.
In particular, if $\hat{A}_{\vb*{\varphi}}^\dag$ has a more simplified expression, such as the displacement operator of coherent state, we do not have to perform Taylor series expansion on it.
A simpler approach is to directly decompose the creation operators of this simplified form.

It should be emphasized that the equivalence in the number of basis functions between sets $\{\psi_n(\omega)\}_{n=1}^N$ and $\{\varphi_n(\omega)\}_{n=1}^N$ is imposed solely to simplify the summation notation;
this is not a strict requirement in practice.

When interference occurs, coherent interference emerges exclusively between $\psi_n(\omega)$ within the same mode, while no cross-mode coherent interference takes place.
The resultant state manifests as a product state of independently interfered modes.

\subsection{Several HOM Interference Examples}
% \begin{figure}[h]
% 	\centering
% 	\includegraphics[width=0.48\textwidth]{HOM_interference_scheme.pdf}
% 	\caption{A scheme of HOM interference.
% 	The incoming photons via paths $a$ and $b$ interfere on a 50:50 BS, and then be detected by detectors $a$ and $b$ with detection efficiency $\eta_1$ and $\eta_2$, respectively.
% 	The dark counts of both detectors are zero.
% 	A coincident count occurs when both detectors click.}
% 	\label{fig: HOM interference}
% \end{figure}
We make three assumptions about the experimental structure of HOM interference:
(1) the transformation on the creation operators by BS \cite{Scully}
\begin{align}
	\hat{a}^\dag &\xrightarrow{\text{BS}} \frac{\hat{a}^\dag+\hat{b}^{\dag}}{\sqrt{2}},\\
	\hat{b}^\dag &\xrightarrow{\text{BS}} \frac{\hat{a}^\dag-\hat{b}^{\dag}}{\sqrt{2}}
\end{align}
acts exclusively on the spatial (here path-encoded) modes, leaving other degrees of freedom (TMs, polarization modes, etc.) unaffected \cite{Tichye}.
\textcolor{blue}{Therefore, the polarization compatibility of our theory is natural};
(2) the dark counts of the detectors $a$ and $b$, with detection efficiency $\eta_1$ and $\eta_2$ respectively, are both zero;
(3) all incident photons share identical polarization.

Additionally, the last assumption is included only for simplicity in the following discussion.
If polarization information is needed, a polarization degree of freedom independent of the temporal mode can be attached to the creation operator.
A specific polarization-involved interference result can be seen in Sec. \ref{sec: asymmetric MDI}.

\subsubsection{Single-Photon Single-Mode Interference}
As a simple introduction, we consider the case that there is only a single photon in TM $\psi$ in path $a$, while another single photon in TM $\varphi$ in path $b$.
Then we can decompose the creation operation in path $b$:
\begin{align}
	\hat{b}_\varphi^\dag &= c\hat{b}_\psi^\dag+\sqrt{1-\abs{c}^2}\hat{b}_\bot^\dag,
\end{align}
where $c=\mel{0}{\hat{b}_\psi\hat{b}_\varphi^\dag}{0}$.
The input state can be written as
\begin{align}
	\ket{\Psi_{\text{in}}} &= \ket{1,\psi}_a\ket{1,\varphi}_b=\hat{a}_\psi^\dag\hat{b}_\varphi^\dag\ket{0}\notag\\
	&= \hat{a}_\psi^\dag(c\hat{b}_\psi^\dag+\sqrt{1-\abs{c}^2}\hat{b}_\bot^\dag)\ket{0}.
\end{align}
Then the output state after BS is
\begin{align}
	\ket{\Psi_{\text{out}}} =& \left[c\frac{(\hat{a}_\psi^\dag)^2-(\hat{b}_\psi^\dag)^2}{2}+\sqrt{1-\abs{c}^2}\right.\notag\\
	&\times\left.\frac{\hat{a}_\psi^\dag+\hat{b}_\psi^\dag}{\sqrt{2}}\frac{\hat{a}_\bot^\dag-\hat{b}_\bot^\dag}{\sqrt{2}}\right]\ket{0}.
\end{align}
The state containing coincident count events is
\begin{align}
	\ket{\Psi_{\text{coin}}} &= \frac{\sqrt{1-\abs{c}^2}}{2}\qty(\hat{a}_\bot^\dag\hat{b}_\psi^\dag-\hat{a}_\psi^\dag\hat{b}_\bot^\dag)\ket{0},
\end{align}
and the corresponding coincidence probability is
\begin{align}
	&P(\text{coin})=\eta_1\eta_2\braket{\Psi_{\text{coin}}}{\Psi_{\text{coin}}}=\frac{\eta_1\eta_2}{2}(1-\abs{c}^2)\notag\\
	=& \frac{\eta_1\eta_2}{2}\qty(1-\int\dd\omega_1\psi^*(\omega_1)\varphi(\omega_1)\int\dd\omega_2\psi(\omega_2)\varphi^*(\omega_2)).
\end{align}
If $\eta_1=\eta_2=100\%$, this result is consistent with the result in Ref. \cite{Agata,Young}.

\subsubsection{Multi-photon Single-Mode Interference}
Now we consider the coherent sources $\alpha$ in mode $\psi$ and $\beta$ in mode $\varphi$, and the corresponding displacement operator is denoted as $\hat{D}_{a,\psi}(\alpha)$ and $\hat{D}_{b,\varphi}(\beta)$.
Similarly, we decompose the initial state in path $b$:
\begin{align}
	&\ket{\beta,\varphi}_b=\hat{D}_{b,\varphi}(\beta)\ket{0}=\exp(\beta\hat{b}_\varphi^\dag-\beta^*\hat{b}_\varphi)\ket{0}\notag\\
	=& \exp\qty[(\beta c\hat{b}_\psi^\dag-\text{H.c.})+(\beta\sqrt{1-\abs{c}^2}\hat{b}_\bot^\dag-\text{H.c.})]\ket{0}\notag\\
	=& \hat{D}_{b,\psi}(\beta c)\hat{D}_{b,\bot}(\beta\sqrt{1-\abs{c}^2})\ket{0}\notag\\
	=& \ket{\beta c,\psi}_b\ket{\beta(1-\abs{c}^2)^{-1/2},\bot}_b,
	\label{eq: Multimode of the coherent state}
\end{align}
where Baker-Campbell-Hausdorff formula \cite{Scully} and Eq.\eqref{eq: commutativity of different TMs} has been applied in the third line.
\textcolor{blue}{For an intuitive description of asymmetry reflected by $c$, we assume Gaussian-shaped $\psi(\omega)$ and $\varphi(\omega)$:
\begin{align}
	\psi(\omega) &= \frac{1}{\pi^{1/4}\sigma_1^{1/2}}\e^{-(\omega-\omega_1)^2/2\sigma_1^2},\\
	\varphi(\omega) &= \frac{1}{\pi^{1/4}\sigma_2^{1/2}}\e^{-(\omega-\omega_2)^2/2\sigma_2^2},
\end{align}
with temporal delay $\tau$, where $\sigma_i^2$ is twice the variance.
By definition,
\begin{align}
	c &= \int\psi^*(\omega)\varphi(\omega)\e^{\text{i}\omega\tau}\dd{\omega}\notag\\
	&= \exp[-\frac{(\Delta\omega)^2+\sigma_1^2\sigma_2^2\tau^2}{2(\sigma_1^2+\sigma_2^2)}]\exp\qty[\text{i}\frac{\sigma_2^2\omega_1+\sigma_1^2\omega_2}{\sigma_1^2+\sigma_2^2}\tau],
\end{align}
where $\Delta\omega=\omega_2-\omega_1$.
As the difference between two central frequency and the delay increase, the coefficient $c$ reflecting overlap will decrease exponentially.
For given and fixed non-zero $\Delta\omega$ and $\tau$, it can be proved that $\abs{c}\leqslant\exp(-\Delta\omega\tau/2)$ with equality if and only if $\sigma_1=\sigma_2=\sqrt{\Delta\omega/\tau}$.
In extreme asymmetric case, $c$ will severely degrade to zero, where no interference occurs.}

The output state can be written as
\begin{align}
	\ket{\Psi_\text{out}} =& \ket{\frac{\alpha+\beta c}{\sqrt{2}},\psi}_a\ket{\frac{\alpha-\beta c}{\sqrt{2}},\psi}_b\notag\\
	&\bigotimes\ket{\beta\sqrt{\frac{1-\abs{c}^2}{2}},\bot}_a\ket{-\beta\sqrt{\frac{1-\abs{c}^2}{2}},\bot}_b
\end{align}
Let
\begin{align}
	\mu_a &= \frac{\abs{\alpha+\beta c}^2}{2}+\abs{\beta}^2\frac{1-\abs{c}^2}{2},\notag\\
	\mu_b &= \frac{\abs{\alpha-\beta c}^2}{2}+\abs{\beta}^2\frac{1-\abs{c}^2}{2}.
\end{align}
Because mode $\psi$ is independent of mode $\bot$, based on Eq.\eqref{eq: Poissonian generating function} and Eq.\eqref{eq: total generating function},
the probability that there exist $n_i$ (i=$a$,$b$) photons in path $i$ is
\begin{align}
	P_i(n_i) &=\frac{1}{n_i!}\eval{\pdv[n_i]{z}\text{e}^{(z-1)\mu_i}}_{z=0}=\text{e}^{-\mu_i}\frac{\mu_i^{n_i}}{n_i!}.
\end{align}

When $n_i$ photons are incident on the detector, the click probability \cite{Rarity} is $P_{n_i}(\text{click})=1-(1-\eta)^{n_i}$.
So the coincidence probability is
\begin{align}
	P(\text{coin}) =& \sum_{m=0}^\infty P_a(m)[1-(1-\eta_1)^m]\notag\\
	&\times\sum_{n=0}^\infty P_b(n)[1-(1-\eta_2)^n]\notag\\
	=& \qty(1-\text{e}^{-\eta_1\mu_a})\qty(1-\text{e}^{-\eta_2\mu_b}).
\end{align}
If the coherent states input in two paths are completely indistinguishable, i.e. $\abs{c}=1$, then $\mu_a=\abs{\alpha+\beta}^2/2$, $\mu_b=\abs{\alpha-\beta}^2/2$, and thereby
\begin{align}
	\eval{P(\text{coin})}_{\abs{c}=1} &= (1-\text{e}^{-\eta_1\abs{\alpha+\beta}^2/2})(1-\text{e}^{-\eta_2\abs{\alpha-\beta}^2/2}),
\end{align}
where we attribute the phase of $c$ to $\beta$.
If they are completely distinguishable, i.e. $c=0$, then $\mu_a=\mu_b=(\abs{\alpha}^2+\abs{\beta}^2)/2$, and thereby
\begin{align}
	\eval{P(\text{coin})}_{c=0} =& \qty[1-\text{e}^{-\eta_1(\abs{\alpha}^2+\abs{\beta}^2)/2}]\notag\\
	&\times\qty[1-\text{e}^{-\eta_2(\abs{\alpha}^2+\abs{\beta}^2)/2}].
\end{align}
These two results are consistent with the result in Ref. \cite{Chen}.

\subsubsection{Single-Photon Multimode Interference}
We consider the interference of a pair of entangled photons generated in the SPDC process, which is detailed in Appendix \ref{appendix: SPDC}.
Given by Ref. \cite{Agata,Jin}, the initial state can be written as
\begin{align}
	\ket{\Psi_{\text{in}}} &= \int\dd\omega_s\int\dd\omega_i f(\omega_s,\omega_i)\hat{a}_s^\dag(\omega_s)\hat{a}_i^\dag(\omega_i)\ket{0}\notag\\
	&= \sum_{m=1}^N\sqrt{\lambda_m}\hat{a}_{\psi,m}^\dag\hat{b}_{\varphi,m}^\dag\ket{0},
\end{align}
where Eq.\eqref{eq: Schmidt decomposition}, Eq.\eqref{eq: signal creation} and Eq.\eqref{eq: idler creation} has been applied in the last line.
The next step is to introduce a unitary transformation matrix $\vb*{c}$ to decomposition $\hat{b}_{\varphi,m}^\dag$:
\begin{align}
	\hat{b}_{\varphi,m}^\dag &= \sum_{n=1}^Nc_{mn}\hat{b}_{\psi,n}^\dag,
\end{align}
where $c_{mn}=\mel{0}{\hat{b}_{\psi,n}\hat{b}_{\varphi,m}^\dag}{0}$.
At this time, the initial state can be rewritten as
\begin{align}
	\ket{\Psi_{\text{in}}} &= \sum_{m,n=1}^Nc_{mn}\sqrt{\lambda_m}\hat{a}_{\psi,m}^\dag\hat{b}_{\psi,n}^\dag\ket{0}.
\end{align}
Note that now $\hat{a}_{\psi,n}^\dag$ and $\hat{b}_{\psi,n}^\dag$ are in the same TM, for simplicity, we omit the subscript $\psi$.
After BS, the final state and the containing coincident count events are
\begin{align}
	\ket{\Psi_{\text{out}}} &= \sum_{m,n=1}^Nc_{mn}\sqrt{\lambda_m}\frac{\hat{a}_m^\dag+\hat{b}_m^\dag}{\sqrt{2}}\frac{\hat{a}_n^\dag-\hat{b}_n^\dag}{\sqrt{2}}\ket{0},\\
	\ket{\Psi_{\text{coin}}} &= \frac{1}{2}\sum_{m,n=1}^Nc_{mn}\sqrt{\lambda_m}(\hat{a}_n^\dag\hat{b}_m^\dag-\hat{a}_m^\dag\hat{b}_n^\dag)\ket{0}.
\end{align}
Therefore, for ideal detectors ($\eta_1=\eta_2=100\%$), the coincidence probability is 
\begin{align}
	&P(\text{coin})=\braket{\Psi_{\text{coin}}}{\Psi_{\text{out}}}\notag\\
	=& \frac{1}{2}\sum_{m,n,k,l=1}^Nc_{mn}^*c_{kl}\sqrt{\lambda_m\lambda_k}(\delta_{mk}\delta_{nl}-\delta_{nk}\delta_{ml})\notag\\
	=& \frac{1}{2}\qty(\sum_{m=1}^N\lambda_m\sum_{n=1}^N\abs{c_{mn}}^2-\sum_{m,n=1}^Nc_{mn}^*c_{nm}\sqrt{\lambda_m\lambda_n})\notag\\
	=& \frac{1}{2}-\frac{1}{2}\sum_{m,n=1}^N\sqrt{\lambda_m\lambda_n}\int\dd\omega_1\psi_n(\omega_1)\varphi_m^*(\omega_1)\notag\\
	&\times\int\dd\omega_2\psi_m^*(\omega_2)\varphi_n(\omega_2)\notag\\
	=& \frac{1}{2}-\frac{1}{2}\int\dd\omega_1\int\dd\omega_2\sum_{m=1}^N\sqrt{\lambda_m}\psi_m^*(\omega_2)\varphi_m^*(\omega_1)\notag\\
	&\times\sum_{n=1}^N\sqrt{\lambda_n}\psi_n(\omega_1)\varphi_n(\omega_2)\notag\\
	=& \frac{1}{2}-\frac{1}{2}\int\dd\omega_1\int\dd\omega_2f^*(\omega_2,\omega_1)f(\omega_1,\omega_2),
\end{align}
where the unitary property of matrix $\vb*{c}$ has been applied in the third line.
This result is consistent with the result in Ref. \cite{Agata,Jin}.

\section{MDI-QKD with Asymmetric Sources}
\label{sec: asymmetric MDI}
In this section, we formally analyze the asymmetric MDI-QKD protocol where the two communication parties use a WCP source and a SPDC source, respectively.
% \textcolor{blue}{It should be emphasized that, this is a specific example for polarization-encoding MDI-QKD, which verifies the polarization compatibility of our theory.}
First, we introduce an experimental setup for implementation.
Secondly, we decompose the coherent state in the TMs of the signal photons from certain SPDC source, and calculate the interference result.
Thirdly, we consider the finite-size effect, which introduce statistic fluctuations, leading to lower key rate and necessarily more optimization parameters.
Finally, we present a comparative analysis of the key rate simulations for symmetric WCP sources, symmetric SPDC sources, and asymmetric sources mentioned above.
\begin{figure}[h]
	\centering
	\includegraphics[width=0.48\textwidth]{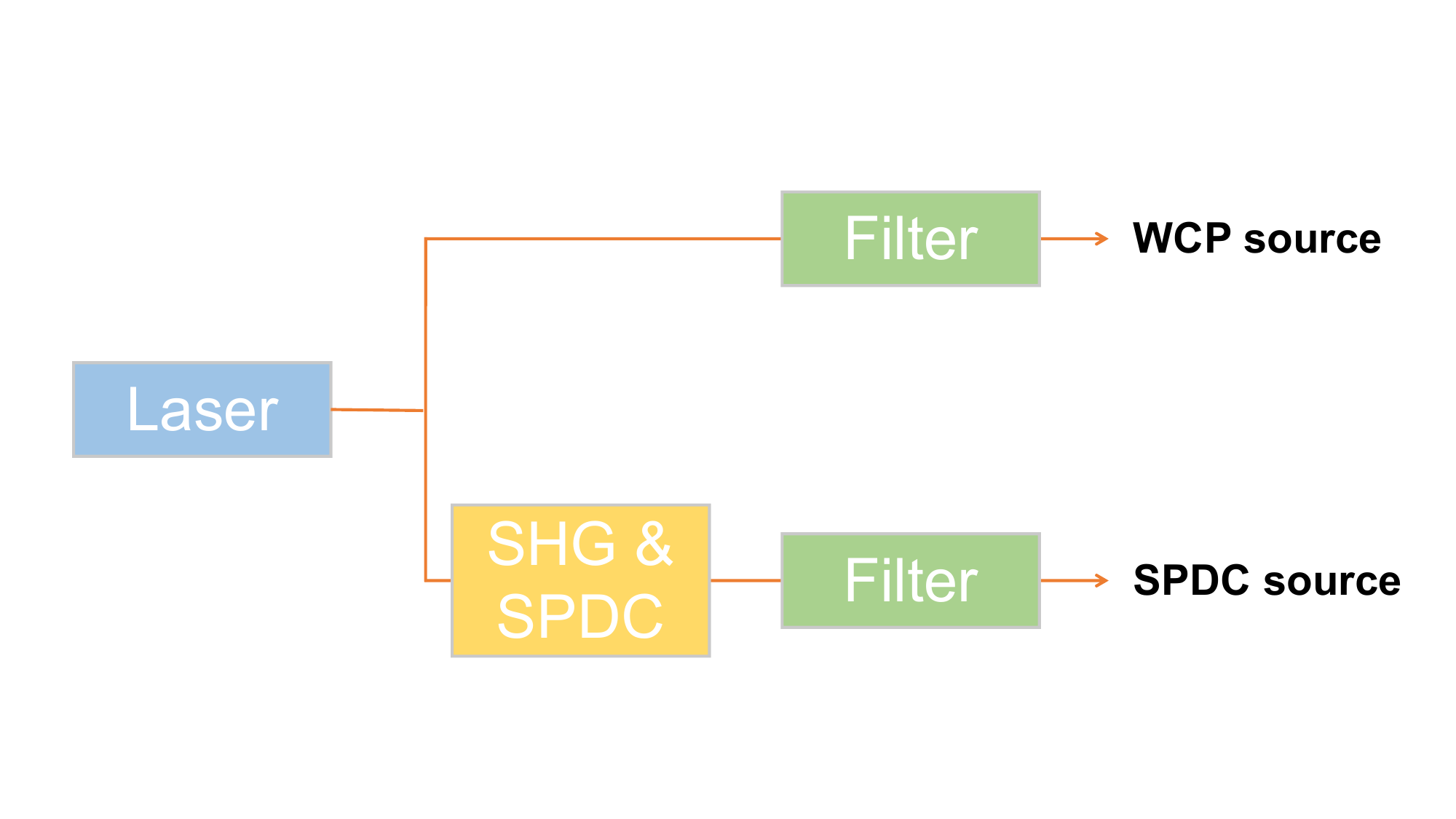}
	\caption{A simplified scheme of experimental setup.
	The emitted pulse laser beam is split into two components: one is utilized as the WCP source after a Gaussian filter,
	while the other is employed as the pump pulse in the SPDC process.
	SHG means second harmonic generation process.
	These two beams are then subjected to MDI-QKD protocol implementation, with interference occurring at a 50:50 BS.}
	\label{fig: experimental setup}
\end{figure}

\subsection{Expected Experimental Structure}
\label{subsec: Experimental Setup}
According to the experimental setup in Ref. \cite{TakesueTeleportation}, we give a simplified scheme of the structure in Fig. \ref{fig: experimental setup}.
To facilitate simulation, we assume that the probability normalized laser waveform of in the frequency domain is independent of the laser intensity.
Therefore, the waveform of the WCP source and the pump pulse in the SPDC process are the same, which is assumed to be Gaussian-shaped:
\begin{align}
	\alpha_0(\omega) &= \frac{1}{\pi^{1/4}\sigma^{1/2}}\exp\qty[-\frac{(\omega-\bar{\omega})^2}{2\sigma^2}].
\end{align}
The Gaussian filter is used to suppress noise and obtain higher spectral purity \cite{Zhong}, whose mathematical form is
\begin{align}
	F_n(\omega) &= \exp[-2^{2n-1}\ln 2\qty(\frac{\omega-\omega_0}{\omega_{\text{FWHM}}})^{2n}],
\end{align}
where $n$ is a positive integer and $\omega_{\text{FWHM}}$ is the full width at half power, i.e. $\abs{F_n(\pm\omega_{\text{FWHM}}/2)}^2=1/2$.
The final waveform of WCP source and SPDC source are $\alpha(\omega)=\alpha_0(\omega)F(\omega)$ and $g(\omega_s,\omega_i)=f(\omega_s,\omega_i)F_s(\omega_s)F_i(\omega_i)$, respectively.

\subsection{Interference of WCP Source and SPDC Source}
As a specific example of the MDI-QKD with asymmetric sources, we let the signal photons from an SPDC source and coherent photons interfere on BS.
It is the interference of single-mode photons and multimode photons.
We choose the TMs of signal photons showed in Eq.\eqref{eq: JSA} as a set of orthonormal function bases.
From Eq.\eqref{eq: Multimode decomposition} and Eq.\eqref{eq: Multimode of the coherent state}, the initial state of WCP source can be decomposed into
\begin{align}
	\ket{\alpha}_b &= \bigotimes_{i=1}^N\ket{\alpha c_i},
	\label{eq: decomposed coherent state}
\end{align}
where
\begin{align}
	c_i=\mel{0}{\hat{b}_{\psi,i}\hat{b}_{\alpha}^\dag}{0}=\int\dd\omega\psi_i^*(\omega)\alpha(\omega).
\end{align}
We assume that the dark count rate $d_S$ and detection efficiency $\eta_S$ of the detectors at the measurement party (Charlie) of the MDI-QKD protocol are also identical.
In such a case, we can attribute the detection efficiency to the channel loss \cite{WangMDI},
and the PND of $n$ photons and the probability of $\vb*{k}$ photons after the channel are
\begin{align}
	P_\mu(n) &= \frac{1}{n!}\eval{\pdv[n]{z}\prod_{i=1}^N\text{e}^{(z-1)\abs{\alpha c_i}^2}}_{z=0}=\text{e}^{-\mu}\frac{\mu^n}{n!},\\
	f_\mu(\vb*{k}) &= \sum_{\vb*{m}\geqslant\vb*{k}}\prod_{i=1}^N\begin{pmatrix}
		m_i\\ k_i
	\end{pmatrix}\eta_S^{k_i}(1-\eta_S)^{m_i-k_i}P_i(m_i)\notag\\
	&= \text{e}^{-\eta_S\mu}\prod_{i=1}^N\frac{(\eta_S\mu_i)^{k_i}}{k_i!},
\end{align}
where $\mu_i=\abs{\alpha c_i}^2$ and $\mu=\abs{\alpha}^2=\sum_{i=1}^N\mu_i$.
See appendix \ref{appendix: multimode of SPDC} for the detailed derivation process of these two results above.

\subsection{Optimization with Finite-Size Effect}
We use the general framework for multimode-photon interference proposed in Ref. \cite{Zhan},
and apply the three-intensity decoy-state method \cite{LoDecoy, WangDecoy} to get the overall gain and quantum bit error rate (QBER) in rectilinear (or $Z$) and diagonal (or $X$) bases, respectively.
In practice, the number of communication rounds $N$ are experimentally finite, then the finite-size effect, essentially the statistic fluctuation, should be considered.

First, the product of the calculated total data size $N$ and the overall gain $Q$ corresponds to the counts of successful events measured in the experiment;
and the product of $N$ and QBER $E$ corresponds to the number of error bits.
We denote $\mu,\nu$ as the intensities from two parties, Alice and Bob. 
Because the formulas \cite{MDI}
\begin{align}
	Q_{\mu\nu} &= \sum_{m,n=0}^\infty P_\mu(m)P_\nu(n)Y_{mn},\\
	E_{\mu\nu}Q_{\mu\nu} &= \sum_{m,n=0}^\infty P_\mu(m)P_\nu(n)Y_{mn}e_{mn}
\end{align}
are applicable only to asymptotic cases, that is, the size of data is infinite.
Therefore, we must estimate the lower and upper bound of the expected value $E(X)$ with the measured value $X$ \cite{Jiang}:
\begin{align}
	E^L(X,\xi)=\frac{X}{1+\delta_1(X,\xi)},\\
	E^U(X,\xi)=\frac{X}{1-\delta_2(X,\xi)},
\end{align}
where $\delta_1,\delta_2$ satisfy the following equations:
\begin{align}
	\delta_1-(1+\delta_1)\qty[\ln(1+\delta_1)+\frac{\ln\xi}{X}]=0,\\
	\delta_2+(1-\delta_2)\qty[\ln(1-\delta_2)+\frac{\ln\xi}{X}]=0,
\end{align}
where $\xi$ is the failure probability.

Consider that $Y_{mn},e_{mn}<1$, we set a cutoff number of photons $N_{\text{max}}$, then the linear programming problem is
\begin{align}
	0<Q_{\mu\nu}&-\sum_{m,n=0}^{N_{\text{max}}}P_\mu(m)P_\nu(n)Y_{mn}\notag\\
	&<1-\sum_{m,n=0}^{N_{\text{max}}}P_\mu(m)P_\nu(n),\\
	0<E_{\mu\nu}Q_{\mu\nu}&-\sum_{m,n=0}^{N_{\text{max}}}P_\mu(m)P_\nu(n)Y_{mn}e_{mn}\notag\\
	&<1-\sum_{m,n=0}^{N_{\text{max}}}P_\mu(m)P_\nu(n).
\end{align}

After we obtain the lower bound of the expected value of $Y_{11}$ and the upper bound of the expected value of $Y_{11}e_{11}$,
it should be noted that the experimentally measured values still exhibit statistical fluctuations relative to the expected values.
So we use the Chernoff bound \cite{Jiang} to estimate a lower $Y_{11}$ and a higher $Y_{11}e_{11}$:
\begin{align}
	O^U(Y,\xi)=[1+\delta_1'(Y,\xi)]Y,\\
	O^L(Y,\xi)=[1-\delta_2'(Y,\xi)]Y,
\end{align}
where $O$ is the observed value, $Y=E(O)$ is the expected value, and $\delta_1',\delta_2'$ satisfy the following equations:
\begin{align}
	\delta_1'-(1+\delta_1')\ln(1+\delta_1')-\frac{\ln\xi}{Y}=0,\\
	\delta_2'+(1-\delta_2')\ln(1-\delta_2')+\frac{\ln\xi}{Y}=0.
\end{align}
Finally, we obtain the key rate as conservative as possible in the experiment.

\subsection{Simulation Results}
\textcolor{blue}{In this subsection, we give a clear comparison curve of key rates with different sources.
We denote W as WCP source and S as SPDC source.
To determine the proportion of each mode in SPDC source, we consider the down-conversion process 532 nm $\to$ 1064 nm $+$ 1064 nm.
For simplicity and intuitiveness, we assume phase matching.
The standard deviation of the Gaussian-shaped pump pulse is 5 ps, and the full width at half maximum (FWHM) of the filter $\omega_{\text{FWHM}}=600$ GHz.
The frequency spectrum and the corresponding spectrum of the first and second modes has been shown in Fig. \ref{fig: JSA}.}

\textcolor{blue}{In simulation, we set $\eta_S=60\%$, $d_S=10^{-6}$ for Charlie, and $\eta_I=90\%$, $d_I=10^{-6}$ for both Alice and Bob;
the loss per kilometer is $\alpha=0.2$ dB/km;
the misalignment error is $e_d=1.5\%$;
the failure probability when estimating Chernoff bound is $\xi=10^{-7}$;
the cutoff number of photons in linear programming is $N_{\max}=6$.}

\begin{figure}[h]
	\centering
	\includegraphics[width=0.48\textwidth]{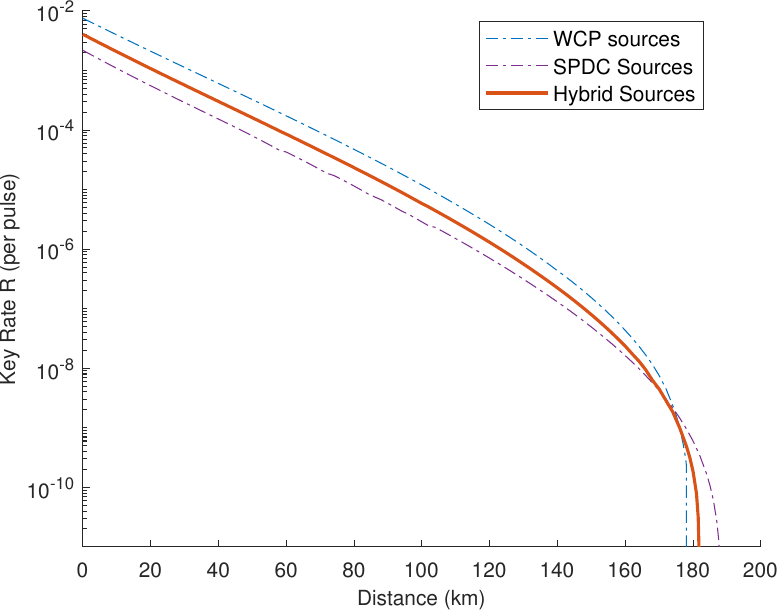}
	\caption{This figure shows the key rates variation with distance between Alice and Bob.
		The blue dashdot line, purple dashdot line and the red solid line represent W-W, S-S and W-S communications, respectively.
		We set the data size $N_{\text{tot}}=10^{12}$.}
	\label{fig: comparison curve (loss)}
\end{figure}

\begin{figure}[h]
	\centering
	\includegraphics[width=0.48\textwidth]{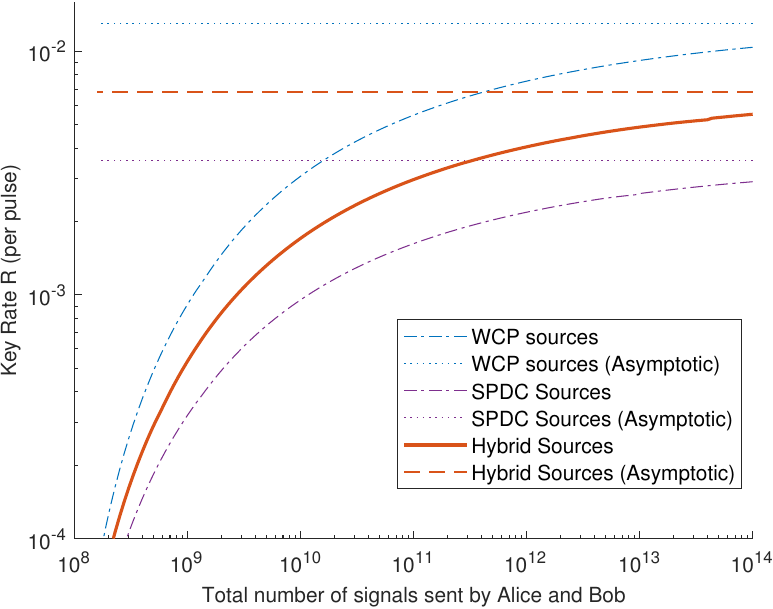}
	\caption{This figure shows the key rates variation with total number of signals sent by Alice and Bob, ranging from $10^8$ to $10^{14}$.
		The blue dashdot line, purple dashdot line and the red solid line represent W-W, S-S and W-S communications, respectively.
		And the blue dotted line, purple dotted line and the red dashed line represent the corresponding asymptotic key rates.
		We set the detection efficiency $\eta_S=60\%$.}
	\label{fig: comparison curve (size)}
\end{figure}

\textcolor{blue}{We optimize the intensities of decoy and signal states $\nu,\mu$, respectively,
and the four probabilities that Alice and Bob choose $\mu$ or $\nu$ in $Z$ or $X$ basis.
The parameters to be optimized amounts to 6 in symmetric W-W and S-S communications, and 12 in W-S communication.
We apply particle swarm optimization (PSO) method to obtain the initial (i.e., lossless) global optimal point.
As channel loss increases, an improved search algorithm is applied, and each point to be optimized starts with the previous optimal value.}
Based on Gottesman-Lo-Lütkenhaus-Preskill (GLLP) formula \cite{GLLP}, the key rate formula \cite{Jiang} we utilize is
\begin{align}
	R\geqslant& p_{\mu_A}^Zp_{\mu_B}^Z\{P_{\mu_A}(1)P_{\mu_B}(1)Y_{11}^Z[1-H_2(e_{11}^X)]\notag\\
	&-Q_{\mu_A\mu_B}^Zf(E_{\mu_A\mu_B}^Z)H(E_{\mu_A,\mu_B}^Z)\},
\end{align}
where $p_{\mu_A}^Z$ and $p_{\mu_B}^Z$ denote the probability that Alice and Bob choose signal state in $Z$ basis, respectively;
$H_2(x)=-x\log_2x-(1-x)\log(1-x)$ is the binary Shannon entropy;
$Y_{11}^Z$ and $e_{11}^X$ denote the single-photon yield in $Z$ basis and QBER in $X$ basis, which can be seen as the phase error in $Z$ basis;
$Q_{\mu_A\mu_B}^Z$ and $E_{\mu_A\mu_B}^Z$ denote the overall gain and QBER in $Z$ basis;
$f(E_{\mu_A\mu_B}^Z)\approx1.16$ is an inefficiency function for the error correction process \cite{MDI}.
The simulation results are in Fig. \ref{fig: comparison curve (loss)} and Fig. \ref{fig: comparison curve (size)}.

Fig. \ref{fig: comparison curve (loss)} demonstrates that at low loss, the single-photon yield $Y_{11}$ plays a dominant role in GLLP formula.
Since the heralded single-photon source (here, an SPDC source) requires joint response monitoring with local detectors, its single-photon yield in $Z$ basis $Y_{11}^Z$ reduces.
Consequently, the key rates of W-W, W-S and S-S communications decrease sequentially.
As the loss increases, the number of successful events decreases, and the effect of QBER becomes gradually significant.
At this point, if local detector triggers, the probability of vacuum emitted by signal or decoy state equals to $d_I$, which is tiny, thus lowering the QBER.
Therefore, at a long distance, the key rate result is exactly the opposite.
Overall, W-S communication combines the features of the WCP source and the SPDC source, which manifests not quite low key rate over the entire selected distance range.

Fig. \ref{fig: comparison curve (size)} demonstrates that the variation of the data size does not affect the relative relationship of the key rates in different communications (at least at low loss).
And the overall growth trends of all key rates are similar, slowing when the data size exceeds $10^{12}$.

\begin{figure}[h]
	\centering
	\includegraphics[width=0.48\textwidth]{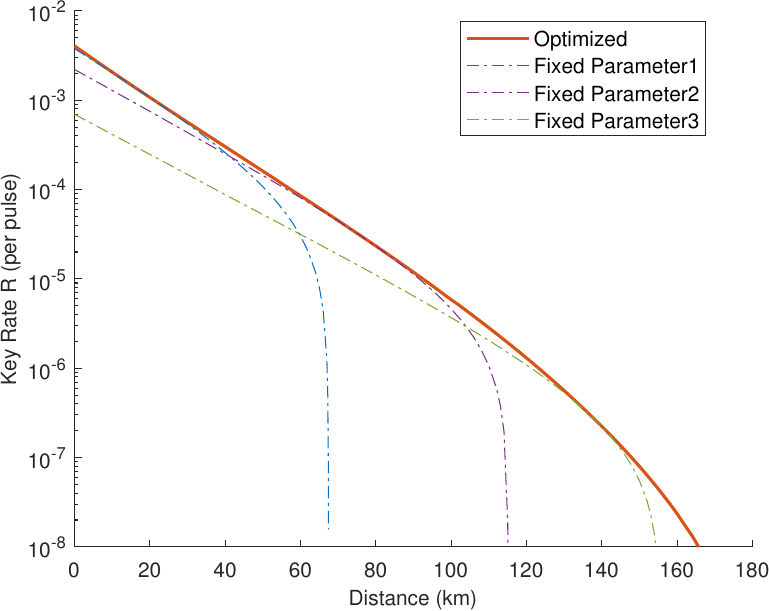}
	\caption{This figure shows the key rates variation with distance from Alice to Bob.
	The red solid line, blue dashdot line, green dashdot line and purple dashdot line represent the key rates optimized and calculated in three arbitrarily given parameters, respectively.
	All simulation parameters remain unchanged.
	The fixed parameters we choose are from the tenth, fortieth and seventieth optimal parameters.}
	\label{fig: comparison curve (fixed params)}
\end{figure}

In addition, we choose the tenth, fortieth and seventieth optimal parameters (see appendix \ref{appendix: fixed parameters}) and fix them at any distance.
That is the reason why the optimal curve and each of the curves with fixed parameters Fig. \ref{fig: comparison curve (fixed params)} have an intersection.
The result shows that, if we conduct the experiment of MDI-QKD with asymmetric sources without this theoretical guidance, the performance at short or long distance becomes significantly worse.
Besides, all the cut-off distances  with fixed parameters fall far short of the optimized cut-off distance.

\textcolor{blue}{At the end of the main text, we consider the practical scene and make a brief discussion.
In principle, any type of data post-processing could not rebuild the detailed original information of two spectra $\psi(\omega)$ and $\varphi(\omega)$.
Actually, $c$ eliminates the details of two spectra, yet all observables relate to $c$ at most.
In other words, it is impossible to deduce the specific waveforms of the frequency spectra from these observables.
In fact, the thing to be concern most is not the spectra themselves, but their differences (or overlaps), that is, the coefficient $c=\int\psi^*(\omega)\varphi(\omega)\dd{\omega}$.}

\textcolor{blue}{Since our expression is general with no assumption, for local test, we can estimate the expected visibility according to our priori knowledge on or parameters of the apparatus in advance, or directly conduct a simple interference experiment.
If it is feasible experimentally, we can go further considering the possible performance in QKD.}

\textcolor{blue}{For practical and large-scale applications, we believe that users usually have no local-test approach.
Therefore, fundamental information such as the trusted ranges of waveform, central wavelength (ideally Gaussian, or typically Lorentzian) and linewidth should be provided by the device suppliers.
Before two users connect, the communication relay estimates the most conservative performance and distributes corresponding extra sources.}

\section{Conclusion}
In summary, we propose a general theory of quantum networks for analyzing and optimizing the interference of any different sources in different temporal modes.
This method would be the theoretical basis for addressing the key rate degradation due to the asymmetry of sources in future applications.

As a practical application, this theory demonstrates the performance of W-W, S-S and W-S communications.
This result demonstrates that the features of the asymmetric sources from both parties are reflected in the key rate.
It means that the asymmetry of sources might not be an obstacle for the implementation of quantum networks.
Furthermore, this theory manifests that for better performance, it may be a possible scheme to modulate two sources to make their TMs match.

\begin{acknowledgments}
	This work has been supported by the National Key Research And Development Program of China (Grant No. 2018YFA0306400), the National Natural Science Foundation of China ( No. 62271463, 62171424, and 62105318), the China Postdoctoral Science Foundation (2022M723064,2021M693098); and the Anhui Initiative in Quantum Information Technologies.
\end{acknowledgments}

% \begin{acknowledgments}
% We wish to acknowledge the support of the author community in using
% REV\TeX{}, offering suggestions and encouragement, testing new versions,
% \dots.
% \end{acknowledgments}

\appendix
\section{Basic Principle of SPDC Source}
\label{appendix: SPDC}
In this section, we will give an introduction to the basic principle of SPDC source and the photon number distribution (PND) of this entanglement source.
Starting form the Maxwell's equations, the effective interaction Hamiltonian of the SPDC process has been given in Ref. \cite{SPDC_Hamiltonians} in detail.
Considering that the effective Hamiltonian of the SPDC process has been studied widely \cite{Grice,Zhong,Zhan,HowColors,ma2007quantum,PhysRevA.61.042304},
here we only care about the specific mathematical form of the effective Hamiltonian:
\begin{align}
	\hat{H}_I &= C\int\dd\omega_s\int\dd\omega_i f(\omega_s,\omega_i)\hat{a}_s^\dag(\omega_s)\hat{a}_i^\dag(\omega_i)+\text{H.c.},
\end{align}
where $C$ is a constant, $\omega_x\ (x=s,i)$ are the frequency of the signal and idler photons respectively, and $\hat{a}_x^\dag(\omega_x)$ are the corresponding creation operators at the frequency $\omega_x$,
which satisfy the commutative relation
\begin{align}
	[\hat{a}_x(\omega_x),\hat{a}_y^\dag(\omega_y')]=\delta_{x,y}\delta(\omega_x-\omega_y').
	\label{eq: commutative relation}
\end{align}

In particular, the joint spectral amplitude (JSA) $f(\omega_s,\omega_i)$ reflect the probability amplitude of this entangled photon pair in frequency, which can be expressed as
\begin{align}
	f(\omega_s,\omega_i)=\alpha(\omega_s+\omega_i)\Phi(\omega_s,\omega_i),
	\label{eq: JSA}
\end{align}
where $\alpha(\omega)$ is the pump envelope function, and $\Phi(\omega_s,\omega_i)$ is the phase-matching function \cite{Grice}.
$\alpha(\omega_s+\omega_i),\Phi(\omega_s,\omega_i)$ reflect the conservation conditions of energy and momentum, respectively.
In a uniformly polarized nonlinear crystal,
\begin{align}
	\Phi(\omega_s,\omega_i)=\text{sinc}\ \frac{\Delta k(\omega_s,\omega_i)L}{2}\exp\qty(\text{i}\frac{\Delta k(\omega_s,\omega_i)L}{2}),
\end{align}
where $\Delta k=k_s(\omega_s)+k_i(\omega_i)-k_p(\omega_p)$.

\begin{figure}[h]
	\centering
	\includegraphics[width=0.48\textwidth]{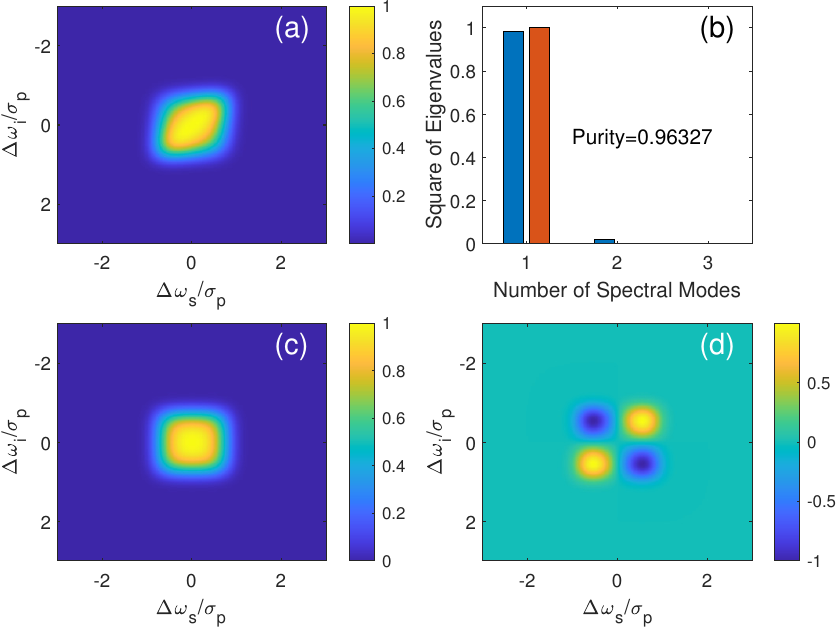}
	\caption{We consider a simple case of phase matching (i.e. $\Delta k=0$) and use Gaussian filters $F(\omega)$ to suppress noise,
	that is, the JSA is now $f(\omega_s,\omega_i)=\alpha(\omega_s+\omega_i)$\protect\newline$F_s(\omega_s)F_i(\omega_s)$.
	The specific expression of Gaussian filters is shown in Sec. \ref{subsec: Experimental Setup}.
	For Gaussian-shaped $\alpha(\omega)$, Fig. (a) shows the JSA relative to its maximum value;
	Fig. (b) shows the \textcolor{blue}{mode proportion comparison between SPDC source (blue bar) and WCP source (orange bar), in which the purity of SPDC source is defined as $\sum_{i=1}^N\lambda_i^2$.
	The "multiple modes" of coherent state originates from the decomposition under the modes of SPDC source};
	Fig. (c) and Fig. (d) show the first and second mode of the JSA to their maximum values after Schmidt decomposition.}
	\label{fig: JSA}
\end{figure}

Now we choose the JSA to be probability normalized, i.e. $\iint\dd\omega_s\dd\omega_i\abs{f(\omega_s,\omega_i)}^2=1$, and then apply the Schmidt decomposition \cite{Law} to Eq.\eqref{eq: JSA}:
\begin{align}
	f(\omega_s,\omega_i)=\sum_{n=1}^\infty\sqrt{\lambda_n}\psi_n(\omega_s)\varphi_n(\omega_i),
	\label{eq: Schmidt decomposition}
\end{align}
where $\{\psi_n(\omega)\}_{n=1}^\infty$ and $\{\varphi_n(\omega)\}_{n=1}^\infty$ are two sets of orthonormal bases.
\textcolor{blue}{Intuitive results of these spectra are shown in Fig. \ref{fig: JSA}.}
In fact, the eigenvalues $\{\lambda_n\}_{n=1}^\infty$, which satisfy $\sum_{n=1}^\infty\lambda_n=1$, decay rapidly with the subscript $n$ for SPDC source, we usually set a cutoff mode number $N$.
This expression reflects the inherent entanglement properties of SPDC sources.
Law et al. \cite{Law} named $\{\psi_n(\omega)\}_{n=1}^N$ and $\{\varphi_n(\omega)\}_{n=1}^N$ as the “temporal modes” (TMs) of signal and idler photons as early as 2000,
after that numerous research on TMs of SPDC sources has emerged \cite{Ansari,Brecht,Raymer,Reddy}.
From this perspective, we can regard JSA as the coherent superposition of these TMs.
For a given subscript $n$, $\psi_n(\omega)$ or $\varphi_n(\omega)$ represent a frequency domain wave function.

From Eq.\eqref{eq: Schmidt decomposition}, we can define effective single-mode field operators by
\begin{align}
	\hat{a}_n^\dag &= \int\dd\omega_s\psi_n(\omega_s)\hat{a}_s^\dag(\omega_s),
	\label{eq: signal creation}\\
	\hat{b}_n^\dag &= \int\dd\omega_i\varphi_n(\omega_i)\hat{a}_i^\dag(\omega_i).
	\label{eq: idler creation}
\end{align}
It is easy to see from Eq.\eqref{eq: commutative relation} that
\begin{align}
	[\hat{a}_m,\hat{a}_n^\dag]=[\hat{b}_m,\hat{b}_n^\dag]=\delta_{mn}.
\end{align}
We assume that the initial state is vacuum state, then the finial state after evolution time $t$ \cite{HowColors} is:
\begin{align}
	\ket{\Psi} &= \exp\qty(\frac{\hat{H}_I}{\text{i}\hbar})\ket{0}\notag\\
	&= \bigotimes_{n=1}^N\exp\qty[\frac{C}{\text{i}\hbar}\qty(\sqrt{\lambda_n}\hat{a}_n^\dag\hat{b}_n^\dag+\text{H.c.})]\ket{0}.
\end{align}
Let $\eta_k=\text{i}C\sqrt{\lambda_k}/\hbar$, and denote $\exp(-\eta_k\hat{a}_k^\dag\hat{b}_k^\dag+\eta_k^*\hat{a}_k\hat{b}_k)$ as $\hat{S}_k(\eta_k)$, then the state of the $k$th mode
\begin{align}
	\ket{\Psi_k} &= \hat{S}_k(\eta_k)\ket{0}=\sum_{n_k=0}^\infty\sqrt{P_k(n_k)}\ket{n_k,n_k}
\end{align}
is a two-mode squeezed state \cite{Scully}, whose probability of $n_k$-photon pair is
\begin{align}
	P_k(n_k) &= \frac{\mu_k^{n_k}}{(1+\mu_k)^{n_k+1}},
	\label{eq: PND in each mode}
\end{align}
where $\mu_k=\sinh^2\abs{\eta_k}=\sinh^2(C\sqrt{\lambda_k}/\hbar)$.

\section{Generating Functions}
For a given discrete probability distribution $P(n)$, its generating function is defined as
\begin{align}
	g(z) &= \sum_{n=0}^\infty P_n(z)z^n.
\end{align}
For example, if discrete random variable $X$ follows the Poisson distribution with parameter $\mu$, then its generating function is
\begin{align}
	g(z) &= \text{e}^{-\mu}\sum_{n=0}^\infty\frac{\mu^n}{n!}z^n=\text{e}^{(z-1)\mu}.
	\label{eq: Poissonian generating function}
\end{align}
The generating function of thermal distribution $P(n)=\mu^n/(1+\mu)^{n+1}$ is
\begin{align}
	g(z) &= \sum_{n=0}^\infty\frac{(\mu z)^n}{(1+\mu)^{n+1}}=\frac{1}{1+\mu-z\mu}.
	\label{eq: thermal generating function}
\end{align}

Generating function possesses a desirable property.
If the probability mass function of random variable $X_i$ (i=1,$\dotsb$,$N$) is $P_{X_i}(n)$, then the distribution of the random variable $Z=\sum_{i=1}^NX_i$ can be calculated as
\begin{align}
	P_Z(n) &= \sum_{\norm{\vb*{n}}_1=n}\prod_{i=1}^NP_{X_i}(n_i)=P_{X_1}(n)*\dotsb*P_{X_N}(n),
\end{align}
where $\norm*{\vb*{n}}_1=\sum_{i=1}^Nn_i$, the symbol “$*$” means convolution.
The generating function can be written concisely:
\begin{align}
	g_Z(z) &= \sum_{n=0}^\infty z^n\sum_{\norm{\vb*{n}}_1=n}\prod_{i=1}^NP_{X_i}(n_i)\notag\\
	&= \sum_{\vb*{n}=0}^\infty\prod_{i=1}^NP_{X_i}(n_i)z^{n_i}=\prod_{i=1}^Ng_{X_i}(z).
	\label{eq: total generating function}
\end{align}
That is, the generating function of the sum is the product of all the individual generating functions.

\section{Calculation of the PND of WCP and SPDC Sources before and after the Channel}
\label{appendix: multimode of SPDC}
We utilize the MDI-QKD framework proposed in Ref. \cite{WangMDI,Zhan} to obtain the overall gain for multimode input.
In the following discussion, we denote the conditional probability $P(A|B)$ as $P_B(A)$.
Assuming a local threshold detector with dark count rate $d_I$ and detection efficiency $\eta_I$, the triggering probability when $k$ photons are incident is \cite{Zhan}
\begin{align}
	P_{k}(\text{trigger}) &= 1-(1-d_I)(1-\eta_I)^k.
\end{align}
It is reasonable to posit that the response of the local detector to $k$ photons states is independent of the intensity $\mu$ of the SPDC source generating such $k$ photons, i.e. $P_k(\text{trigger})=P_{\mu,k}(\text{trigger})$.
We denote $\vb*{k}=(k_1,\dots,k_N)$ as the event that there are $k_n$ photon pairs in the $k_n$th mode from SPDC source, and denote $\norm{\cdot}_1$ as 1-norm of a vector.
From Eq.\eqref{eq: PND in each mode} and the discussion of the independence of different TMs in Subsec. \ref{subsec: independence of different TMs}, the total PND of SPDC source is
\begin{align}
	P_\mu(\vb*{k}) &= P_\mu(k_1,\dotsb,k_N)=\prod_{i=1}^N\frac{\mu_i^{k_i}}{(1+\mu_i)^{1+k_i}}.
	\label{eq: total PND of SPDC}
\end{align}
From Eq.\eqref{eq: thermal generating function}, Eq.\eqref{eq: total generating function} and Eq.\eqref{eq: total PND of SPDC}, the generating function of the probability distribution $\{P_\mu(k)\}_{k=0}^\infty$ that SPDC source emit $k=\norm{\vb*{k}}_1\ (k=1,2,\dotsb)$ photons is
\begin{align}
	g(z) &= \prod_{i=1}^N\frac{1}{1+\mu_i-z\mu_i},
\end{align}
where $\mu_i=\sinh^2(C\sqrt{\lambda_i})$, and thus
\begin{align}
	P_\mu(n)=\frac{1}{n!}\eval{\pdv[n]{g(z)}{z}}_{z=0}=\frac{1}{n!}\pdv[n]{z}\eval{\prod_{i=1}^N\frac{1}{1+\mu_i-z\mu_i}}_{z=0}.
\end{align}
It should be noted that this expression is only suitable for theoretical analysis.
When simulating, numerical convolution
\begin{align}
	P_\mu(n) &= \frac{\mu_1^{n_1}}{(1+\mu_1)^{1+n_1}}*\dotsb*\frac{\mu_N^{n_N}}{(1+\mu_N)^{1+n_N}}
\end{align}
is a more computationally tractable solution.

We assume that the dark count rate $d_S$ and detection efficiency $\eta_S$ of the detectors at the measurement party of the MDI-QKD protocol are also identical.
In such a case, we can attribute the detection efficiency to the channel loss \cite{WangMDI},
and the probability of $\vb*{k}$ photons after the channel is
\begin{align}
	&f_{\mu}(\vb*{k})=\sum_{\vb*{n}\geqslant\vb*{k}}P_{\vb*{n}}(\text{trigger})\notag\\
	&\times\prod_{i=1}^N\begin{pmatrix}
		n_i\\ k_i
	\end{pmatrix}\eta_S^{k_i}(1-\eta_S)^{n_i-k_i}P_{\mu}(\vb*{n})\notag\\
	=& \prod_{i=1}^N\frac{(\eta_S\mu_i)^{k_i}}{k_i!(1+\mu_i)^{1+k_i}}\sum_{\vb*{n}\geqslant0}\qty[1-(1-d_I)(1-\eta_I)^{\sum\limits_{i=1}^N(n_i+k_i)}]\notag\\
	&\times\prod_{i=1}^N\frac{(n_i+k_i)!}{n_i!}\qty[\frac{(1-\eta_S)\mu_i}{1+\mu_i}]^{n_i},
	\label{eq: k-photon probability after channel}
\end{align}
where $\vb*{n}\geqslant\vb*{k}$ means that for each mode $i$, $n_i\geqslant k_i$ holds true.
We denote $C_{N-1}=(1-d_I)(1-\eta_I)^{\sum_{i=1}^Nk_i}(1-\eta_I)^{\sum_{i=1}^{N-1}n_i}$, $x=(1-\eta_S)\mu_N/(1+\mu_N)<1$, $y=(1-\eta_I)x<1$, and consider the sum
\begin{align}
	&\sum_{n_N\geqslant 0}\frac{(n_N+k_N)!}{n_N!}\qty(x^{n_N}-C_{N-1}y^{n_N})\notag\\
	=& \pdv[k_N]{z}\qty(\frac{x_N^{k_N}}{1-x_N}-C_{N-1}\frac{y_N^{k_N}}{1-y_N})\notag\\
	=& \sum_{m=0}^{k_N}\begin{pmatrix}
		k_N\\ m
	\end{pmatrix}(-1)^m\pdv[k_N]{z}\left[(1-x_N)^{m-1}\right.\notag\\
	&-\left.C_{N-1}(1-y_N)^{m-1}\right]\notag\\
	=& k_N!\qty[\frac{1}{(1-x)^{k_N+1}}-\frac{C_{N-1}}{(1-y)^{k_N+1}}]\notag\\
	=& k_N!\left\{\qty(\frac{1+\mu_N}{1+\eta_S\mu_N})^{k_N+1}-C_{N-1}\right.\notag\\
	&\times\left.\qty[\frac{1+\mu_N}{1+(\eta_I+\eta_S-\eta_I\eta_S)\mu_N}]^{k_N+1}\right\}.
	\label{eq: partial summation}
\end{align}
Similarly, each factor $(1-\eta_I)^{n_i}$ in $C_{N-1}$ can provide a sum result
\begin{align}
	k_i!\qty[\frac{1+\mu_i}{1+(\eta_I+\eta_S-\eta_I\eta_S)\mu_i}]^{k_i+1}.
\end{align}
Therefore, the final calculation result of Eq.\eqref{eq: k-photon probability after channel} is
\begin{align}
	f_{\mu,\text{trigger}}(\vb*{k}) =& \prod_{i=1}^N\frac{(\eta_S\mu_i)^{k_i}}{(1+\eta_S\mu_i)^{1+k_i}}-(1-d_I)\notag\\
	&\times\prod_{i=1}^N\frac{[(1-\eta_I)\eta_S\mu_i]^{k_i}}{[1+(\eta_S+\eta_I-\eta_S\eta_I)\mu_i]^{1+k_i}}.
\end{align}

For WCP source decomposed in orthonormal temporal modes in Eq.\eqref{eq: decomposed coherent state}, we adopt the same approach and obtain
\begin{align}
	f(\vb*{k}) &= \sum_{\vb*{n}\geqslant\vb*{k}}\prod_{i=1}^N\begin{pmatrix}
		n_i\\ k_i
	\end{pmatrix}\eta_S^{k_i}(1-\eta_S)^{n_i-k_i}P_{\mu}(\vb*{n})\notag\\
	&= \prod_{i=1}^N\text{e}^{-\mu_i}\frac{(\eta_S\mu_i)^{k_i}}{k_i!}\sum_{n_i=0}^\infty\frac{[(1-\eta_S)\mu_i]^{n_i}}{n_i!}\notag\\
	&= \text{e}^{-\eta_S\mu}\prod_{i=1}^N\frac{(\eta_S\mu_i)^{k_i}}{k_i!},
\end{align}
where $\mu_i=\abs{\alpha c_i}^2$ and $\mu=\abs{\alpha}^2=\sum_{i=1}^N\mu_i$.
The total PND is $P_\mu(n)=\text{e}^{-\mu}\mu^n/n!$.

\textcolor{blue}{\section{Theoretical Analysis of the Visibility with WCP source and SPDC source}}
\textcolor{blue}{Since the expressions of overall gain and QBER are tough for theoretical analysis due to the over-generality in Ref. \cite{WangMDI}, we will apply a more simplified method for given PND.}

\textcolor{blue}{The local detection operator can be described as the projector
    \begin{align}
        \Pi_k &= \op*{\vb*{k}}=\bigotimes_{k_1+\dotsb+k_N=k}\op*{k_i}
    \end{align}
    with corresponding probability $1-(1-d_I)(1-\eta_I)^k$, followed by a partial trace operation on the system of idler photons.
    Note another fact that the quantum channel can be equivalently regarded as a beam splitter \cite{Agata} with transmittance $\eta$, while the remainder of the signal is leaked or consumed.
	In other words, for the export we concern, the transformation on the creation operator is $\hat{a}^\dag\to\sqrt{\eta}\hat{a}^\dag$.
    In this case, for simplicity, we temporarily attribute $\eta$ to $\eta_S$ and omit $\eta_S$.
    After temporal-mode decomposition, the joint state after 50:50 BS in the $i$-th mode can be expressed as
    \begin{align}
        &\ket{\Psi_i}_A\ket{\alpha c_i}_B=\hat{S}_{A,i}(\eta_i)\hat{D}_{B,i}(\alpha c_i)\ket{0}\notag\\
        \xrightarrow{\text{BS}}& \hat{S}_{A,i}\qty(\frac{\eta_i}{\sqrt{2}})\hat{S}_{B,i}\qty(\frac{\eta_i}{\sqrt{2}})\hat{D}_{A,i}\qty(\frac{\alpha}{\sqrt{2}})\hat{D}_{B,i}\qty(-\frac{\alpha}{\sqrt{2}})\ket{0},
    \end{align}
    where $A,B$ represent the sides belong to Alice and Bob, respectively.
	For Side (or Path) $A$, now we execute the projective measurements on idler photons:
    \begin{align}
        \rho_{s,i}(k) &= \Tr_{\text{idler}}\qty[\Pi_k\hat{S}_{A,i}\qty(\frac{\eta_i}{\sqrt{2}})\op*{0}\hat{S}_{A,i}^\dag\qty(\frac{\eta_i}{\sqrt{2}})]\notag\\
		&= \bigotimes_{k_1+\dotsb+k_N=k}P_i(k_i)\op*{k_i},
    \end{align}
    where $P_i(k_i)=\mu_i^{k_i}/(1+\mu_i)^{k_i+1}$ and $\mu_i=\sinh^2\abs{\eta_i/\sqrt{2}}$, then the density operator of the system of signal photons can be written as
    \begin{align}
        \rho_{\text{sigal}} &= \sum_{k=0}^\infty\qty[1-(1-d_I)(1-\eta_I)^k]\rho_{s,i}(k)\notag\\
        &= \bigotimes_{i=1}^N\sum_{k_i=0}^\infty\qty[1-(1-d_I)(1-\eta_I)^{k_i}]P_i(k_i)\op*{k_i},
    \end{align}
    and the total probability of local triggering is
    \begin{align}
        P(\text{trigger}) &= \Tr\rho_{\text{signal}}=1-(1-d_I)\prod_{i=1}^N\frac{1}{1+\eta_I\mu_i}.
    \end{align}}

	\textcolor{blue}{We denote the $i$-th mode of $\rho_{\text{signal}}$ as $\rho_{s,i}$, that is,
    \begin{align}
        &\rho_{s,i}=\sum_{k_i=0}^\infty\qty[1-(1-d_I)(1-\eta_I)^{k_i}]P_i(k_i)\op*{k_i}.
    \end{align}
    Since the majority of non-coincidence is caused by vacuum, we only concern the vacuum probability of the $i$-th mode:
    \begin{align}
        &P_{A,i}(0)=\mel{0}{\hat{D}_{A,i}(\alpha_i)\rho_{s,i}\hat{D}_{A,i}^\dag(\alpha_i)}{0}\notag\\
        =& \frac{1}{1+\mu_i}\qty[\e^{-\abs{\alpha_i}^2/(1+\mu_i)}-(1-d_I)\e^{-(1+\eta_I\mu_i)\abs{\alpha_i}^2/(1+\mu_i)}],
    \end{align}
    where $\alpha_i=\alpha c_i/\sqrt{2}$.
    Therefore, the total probability of vacuum is $P_A(\text{vac})=\prod_{i=1}^NP_{A,i}(0)$.
    By replacing $\alpha,\eta_i$ to $\alpha\sqrt{\eta_S},\eta_i\sqrt{\eta_S}$, the final click probability, $P_A(\text{click})$, of Charlie's detector in Path $A$ can be expressed as
    \begin{align}
        P_A(\text{click}) =& 1-(1-d_S)\prod_{i=1}^N\frac{1}{1+\mu_i}\left[\e^{-\eta_S\abs{\alpha}^2\abs{c_i}^2/2(1+\mu_i)}\right.\notag\\
		&\left.-(1-d_I)\e^{-(1+\eta_I\mu_i)\eta_S\abs{\alpha}^2\abs{c_i}^2/2(1+\mu_i)}\right],
    \end{align}
    where $\mu_i=\sinh^2\abs{\eta_i\sqrt{\eta_S/2}}$.
    Consider the function $f(x)=\e^{-x}-(1-d_I)\e^{-(1+\eta_I\mu)x}\ (x\geqslant0)$.
    Usually $d_I\ll\eta_I\mu_i<1$, and therefore, when $x_0=\ln[(1-d_I)(1+\eta_I\mu)]/(\eta_I\mu)$, function $f(x)$ achieves the maximum.
    Thus, for any temporal mode of WCP source, the click probability for Path $A$ is lower-bounded by
    \begin{align}
        P_A(\text{click}) \geqslant& 1-(1-d_S)\max_i\left\{\frac{\eta_I\mu_i}{(1+\mu_i)(1+\eta_I\mu_i)}\right.\notag\\
        &\left.\qty[\frac{1}{(1-d_I)(1+\eta_I\mu_i)}]^{1/\eta_I\mu_i}\right\}.
    \end{align}
	It is similar for Path $B$.
	With this simplified expression, if we can have a prior knowledge or estimation of $c_i$ (e.g. experimental measurement), we can further judge whether MDI-QKD with these two asymmetric sources is feasible through a pre-determined lower bound of the visibility.}

\section{Fixed Parameters Chosen from the Optimal Parameters}
\label{appendix: fixed parameters}
Here we list the fixed parameters used in Fig. \ref{fig: comparison curve (fixed params)}.
The symbol used in Tab. \ref{tab: fixed parameters} is as follows.
$\nu,\mu$ are respectively the (intensities of) the decoy and signal state;
$P_{\alpha}^\beta\ (\alpha=\mu,\nu,\ \beta=Z,X)$ is the probability that the state $\alpha$ in $\beta$ basis is chosen;
the subscript 1 and 2 respectively represent the parameters of SPDC source and WCP source.

\begin{table*}[htbp]
	\centering
	\caption{The fixed parameters chosen from the optimal parameter of asymmetric sources}
	\label{tab: fixed parameters}
	\begin{tabular*}{\textwidth}{@{\extracolsep{\fill}} ccccccccccccc}
		\hline\hline
		& $\nu_1$ & $\mu_1$ & $P_{\nu_1}^Z$ & $P_{\mu_1}^Z$ & $P_{\nu_1}^X$ & $P_{\mu_1}^X$ & $\nu_2$ & $\mu_2$ & $P_{\nu_2}^Z$ & $P_{\mu_2}^Z$ & $P_{\nu_2}^X$ & $P_{\mu_2}^X$\\
		\hline
		\textbf{Parameter 1} & 0.0066 & 0.4225 & 0.0358 & 0.8754 & 0.0788 & 0.0091 & 0.0542 & 0.5774 & 0.0344 & 0.8757 & 0.0500 & 0.0055\\
		\textbf{Parameter 2} & 0.0082 & 0.2671 & 0.0680 & 0.7453 & 0.1612 & 0.0200 & 0.0624 & 0.4287 & 0.0703 & 0.7466 & 0.1035 & 0.0147\\
		\textbf{Parameter 3} & 0.0134 & 0.1588 & 0.1249 & 0.5094 & 0.3124 & 0.0411 & 0.0812 & 0.3583 & 0.1177 & 0.5144 & 0.2014 & 0.0412\\
		\hline\hline
	\end{tabular*}
\end{table*}

\nocite{*}
\bibliographystyle{unsrtnat}
\bibliography{apssamp}% Produces the bibliography via BibTeX.

\end{document}